\documentclass[prb,aps,showpacs,twocolumn]{revtex4}

\usepackage{amsmath}
\usepackage{bm}
\usepackage{graphicx}

\begin{document}

\title{B\'enard--von K\'arm\'an vortex street in an exciton-polariton
superfluid}

\author{Hiroki Saito}
\affiliation{Department of Engineering Science, University of
Electro-Communications, Tokyo 182-8585, Japan}

\author{Tomohiko Aioi}
\affiliation{Department of Engineering Science, University of
Electro-Communications, Tokyo 182-8585, Japan}

\author{Tsuyoshi Kadokura}
\affiliation{Department of Engineering Science, University of
Electro-Communications, Tokyo 182-8585, Japan}

\date{\today}

\begin{abstract}
The dynamics of an exciton--polariton superfluid resonantly injected into
a semiconductor microcavity are investigated numerically.
The results reveal that a B\'enard--von K\'arm\'an vortex street is
generated in the wake behind an obstacle potential, in addition to the
generation of quantized vortex dipoles and dark solitons.
The vortex street is shown to be robust against a disorder potential in a
sample and it can be observed even in time-integrated measurements.
\end{abstract}


\maketitle

\section{Introduction}

A cylindrical obstacle moving in a classical viscous fluid generates a
vortex-antivortex pair in its wake.
At large Reynolds numbers, such vortices develop into a periodic pattern
known as the B\'enard-von K\'arm\'an vortex (BvK)
street.~\cite{Benard,Karman}
Such dynamics are also found in superfluids for which vortex circulation
is quantized and viscosity is absent.
It was numerically demonstrated~\cite{Frisch} that quantized
vortex-antivortex pairs, which we call vortex dipoles, are shed in the
wake of an obstacle moving in a planar superfluid above a critical
velocity.
This theoretical prediction has been realized in Bose--Einstein
condensates (BECs) of atomic gases, in which quantized vortex dipoles have
been created using an moving obstacle potential produced by a laser
beam.~\cite{Inouye,Neely}
BvK vortex streets in superfluids (i.e., periodic and alternating creation
of quantized vortices and antivortices that form a vortex street with a
long lifetime) have also been predicted for atomic BECs,~\cite{Sasaki}
but they have yet to be demonstrated experimentally.

Exciton-polariton superfluid flow past an obstacle has recently been
demonstrated using a semiconductor
microcavity.~\cite{Amo09N,Amo09,Amo11,Nardin,Sanvitto,Grosso}
In these experiments, polaritons with a controlled in-plane momentum are
coherently injected into a sample by a pumping laser, and an obstacle
potential is produced by a defect in the 
microcavity,~\cite{Amo09N,Amo09,Amo11,Nardin} by a continuous-wave (cw)
laser field,~\cite{Sanvitto} or by etching the sample.~\cite{Grosso}
A Cherenkov-like pattern and oblique dark solitons are observed for 
supersonic flow of polariton condensates.~\cite{Amo09,Amo11}
Oblique dark solitons formed behind the obstacle decay into quantized
vortices.~\cite{Grosso}
For subsonic flow, quantized vortex dipoles are produced in the
wake.~\cite{Amo11,Nardin,Sanvitto}
These phenomena have been studied theoretically by several
researchers.~\cite{Carusotto,Wouters,Cancellieri,Pigeon}

A polariton condensate differs from superfluid helium and an atomic gas
BEC in that it is a nonequilibrium open system.
The polaritons have a short lifetime of $\sim$ 10 ps, which is comparable
to the time scale of the relevant dynamics.
In the experiments described in Refs.~\onlinecite{Amo09,Amo11}, polaritons
are constantly replenished by pumping with a cw laser and the system
reaches a nonequilibrium steady state, in which the pumping balances the
loss.
Another difference from atomic systems is that the polariton is a coherent
superposition of a quantum-well exciton and a cavity photon and the 
interparticle interaction originates only from the former; in other words,
a polariton superfluid is a nontrivial two-component system.
Because of these differences, it is by no means obvious that a polariton
superfluid shares the same dynamic phenomena as superfluid helium and
atomic gas BEC.

In this paper, we investigate the dynamics of an
exciton-polariton superfluid passing an obstacle potential and show that a
superfluid BvK vortex street~\cite{Sasaki} emerges in this system, as well
as vortex dipoles and dark solitons.
We also show that the vortex street is not destroyed by a disorder
potential, which is present in a realistic sample.
We propose a measurement method to identify a vortex street by
time-integrated imaging for the present cw pumped system.

This paper is organized as follows.
Section~\ref{s:form} formulates the problem,
Sec.~\ref{s:result} presents the numerical results, and
Sec.~\ref{s:conc} gives the conclusions of the study.

\section{Formulation of the problem}
\label{s:form}

We consider a system of quantum-well excitons and cavity photons in the
mean-field theory.
The mean-field wave functions of excitons $\psi_{\rm X}$ and photons
$\psi_{\rm C}$ are assumed to obey the two-component nonlinear
Schr\"odinger equation in two dimensions,~\cite{Carusotto}
\begin{widetext}
\begin{equation} \label{GP}
i \hbar \frac{\partial}{\partial t} \left( \begin{array}{c} \psi_{\rm X}
\\ \psi_{\rm C} \end{array} \right) = \left[ H_0 +
\left( \begin{array}{cc} g |\psi_{\rm X}|^2 - i \hbar \gamma_{\rm X} / 2 &
0 \\ 0 & V(\bm{r}) - i \hbar \gamma_{\rm C} / 2 \end{array} \right)
\right] \left( \begin{array}{c} \psi_{\rm X} \\ \psi_{\rm C} \end{array}
\right) + \left( \begin{array}{c} 0 \\ \hbar F(\bm{r}) e^{i (\bm{k}_{\rm
p} \cdot \bm{r} - \omega_{\rm p} t)} \end{array}
\right).
\end{equation}
\end{widetext}
The polariton Hamiltonian in Eq.~(\ref{GP}) is given by
\begin{equation} \label{H0}
H_0 = \hbar \left( \begin{array}{cc} \omega_{\rm X}(-i \nabla) &
\Omega_{\rm R} \\ \Omega_{\rm R} & \omega_{\rm C}(-i \nabla)
\end{array} \right),
\end{equation}
where the diagonal elements are the dispersion relations of an exciton and
a photon, and $\Omega_{\rm R}$ is the Rabi frequency of the exciton-photon
coupling.
In Eq.~(\ref{GP}), $g$ is the exciton-exciton interaction coefficient,
$\gamma_{\rm X}$ and $\gamma_{\rm C}$ are respectively the exciton and
photon decay rates, and $V$ is the potential for cavity photons.
The last term on the right-hand side of Eq.~(\ref{GP}) describes the
coherent quasi-resonant pumping of photons by an external laser beam with
a spatial profile $F$, an in-plane wave vector $\bm{k}_{\rm p}$, and a
frequency $\omega_{\rm p}$.

Diagonalizing the noninteracting Hamiltonian in Eq.~(\ref{H0}), we obtain
the eigenfrequencies of the upper and lower free polaritons as
\begin{equation}
\omega_\pm = \frac{\omega_{\rm X} + \omega_{\rm C} \pm \sqrt{(\omega_{\rm
X} - \omega_{\rm C})^2 + 4 \Omega_{\rm R}^2}}{2}.
\end{equation}
The pumping frequency $\omega_{\rm p}$ is close to $\omega_-$ to
resonantly excite the lower polaritons.
We define the detuning as
$\delta = \omega_p - \omega_-(\bm{k}_{\rm p})$.
The exciton is assumed to have the flat dispersion, $\omega_{\rm
X}(\bm{k}) = \omega_{\rm X}^0$, and the dispersion of the cavity photon is
approximated to be $\omega_{\rm C}(\bm{k}) = \omega_{\rm C}^0 + \hbar k^2
/ (2 m_{\rm C})$, where $m_{\rm C}$ is the effective mass of a cavity
photon.
In the following, we assume $\omega_{\rm X}^0 = \omega_{\rm C}^0$.

We numerically solve Eq.~(\ref{GP}) using the pseudospectral
method,~\cite{Recipes} which imposes periodic boundary conditions.
The spatial range in the numerical calculation is taken to be sufficiently
large that the boundary conditions do not affect the results.
The initial state is the vacuum state of excitons and photons, $\psi_{\rm
X} = \psi_{\rm C} = 0$, plus small white noise to break the numerically
exact symmetry.
The results do not qualitatively depend on the detail of the noise.

\section{Numerical results}
\label{s:result}

\begin{figure}[htbp]
\includegraphics[width=8cm]{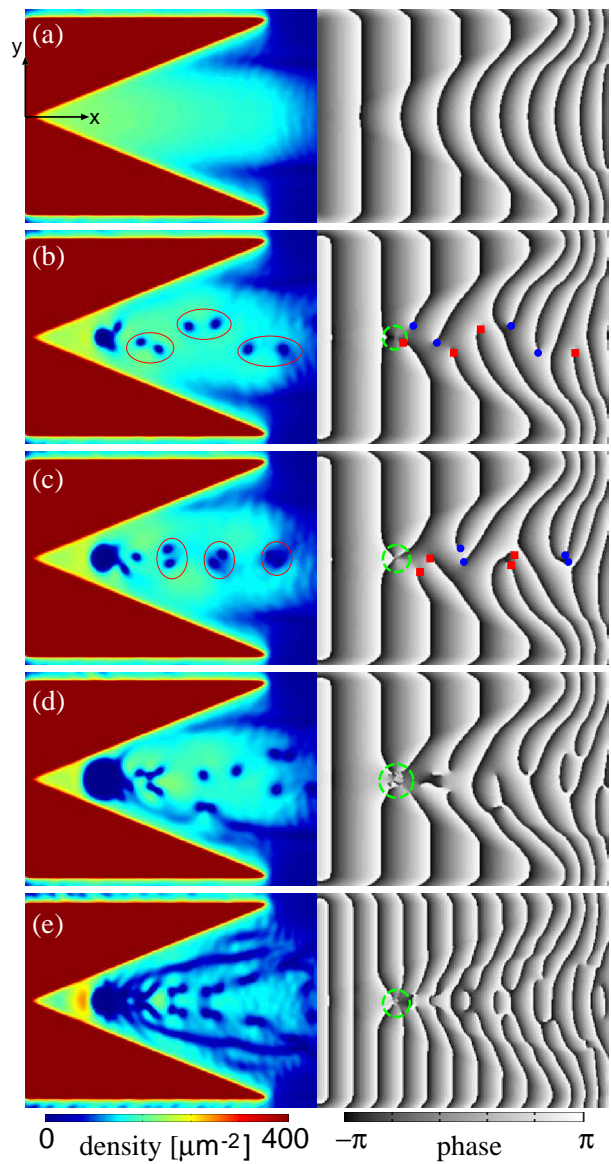}
\caption{
(Color online) Density $|\psi_{\rm C}|^2$ and phase ${\rm
arg}(\psi_{\rm C})$ profiles of the photon wave function at $t = 400$ ps.
(a) $k_{\rm p} = 0.25$ $\mu {\rm m}^{-1}$ without an obstacle potential,
(b) $k_{\rm p} = 0.25$ $\mu {\rm m}^{-1}$ and $d = 2$ $\mu {\rm m}$,
(c) $k_{\rm p} = 0.25$ $\mu {\rm m}^{-1}$ and $d = 3$ $\mu {\rm m}$,
(d) $k_{\rm p} = 0.25$ $\mu {\rm m}^{-1}$ and $d = 5$ $\mu {\rm m}$,
and (e) $k_{\rm p} = 0.5$ $\mu {\rm m}^{-1}$ and $d = 3$ $\mu {\rm m}$.
The solid circles in (b) and (c) respectively indicate vortex dipoles and
co-rotating twin vortices, which are released from the obstacle potential.
The dashed circles indicate the locations of the obstacle potential and the
filled blue circles and red squares indicate clockwise and
counterclockwise vortices, respectively.
The detuning is $\hbar \delta = 0.7$ meV and the lifetime is $\gamma^{-1}
= 30$ ps.
The field of view of each panel is $150 \times 110$ $\mu {\rm m}$ and the
origin is located at the center of the left edge of the panel.
See Supplemental Material for movies of the dynamics in (b)-(e).
}
\label{f:main}
\end{figure}
In the following numerical calculations, we use $m_{\rm C} = 2 \times
10^{-5} m_{\rm e}$, where $m_{\rm e}$ is the electron mass, $\hbar
\Omega_{\rm R} = 5$ meV, and $g = 0.01$ ${\rm meV} \mu{\rm m}^2$, and we
assume $\gamma_{\rm X} = \gamma_{\rm C} \equiv \gamma$.
The pumping function $F(\bm{r})$ is assumed to have the form
\begin{equation} \label{F}
F(\bm{r}) = \left\{ \begin{array}{ll}
F_0 & (\alpha x < |y| < y_{\rm p} \; {\rm and} \; x > 0), \\
0 & ({\rm otherwise}), \end{array} \right.
\end{equation}
which excites the polaritons in the region of the two triangles shown in
Fig.~\ref{f:main} (a).
We restrict the pumping area as in Eq.~(\ref{F}), since if the whole
space is pumped, the phase will be locked and no vortices will be
generated.
Similar pumping schemes have been employed in theoretical~\cite{Pigeon}
and experimental~\cite{Sanvitto} studies.
The parameters in Eq.~(\ref{F}) are taken to be $\hbar F_0 = 38.2$ meV,
$\alpha = 0.4$, and $y_{\rm p} = 50$ $\mu{\rm m}$.
The in-plane wave vector of the pumping beam is $\bm{k}_{\rm p} = k_{\rm
p} \bm{e}_x$, where $\bm{e}_x$ is the unit vector in the $x$ direction.
Polaritons pumped in the triangular areas thus have momentum in the $x$
direction and flow into the region between the triangles, as shown in
Fig.~\ref{f:main} (a).
An obstacle potential is assumed to be a Gaussian potential as
\begin{equation}
V(\bm{r}) = V_0 \exp\left[ -(x - x_0)^2 / d^2 - y^2 / d^2 \right]
\end{equation}
with $V_0 = 38.2$ meV and $x_0 = 40$ $\mu {\rm m}$, which is located
between the triangles.
Thus, polaritons flowing between the triangular areas hit the obstacle
potential and generate a wake in the $x$ direction.
The healing length and the Bogoliubov sound speed near the obstacle
potential are respectively estimated to be $\xi \simeq \hbar / (m_{\rm LP}
g_{\rm LP} n_{\rm LP})^{1/2} \simeq 1.5$ $\mu{\rm m}$ and $v_{\rm s}
\simeq (g_{\rm LP} n_{\rm LP} / m_{\rm LP})^{1/2} \simeq 1.8 \times 10^6$
${\rm m} / {\rm s}$, where $m_{\rm LP}$, $g_{\rm LP}$, and $n_{\rm LP}$
are respectively the mass, the effective coupling
constant,~\cite{Carusotto} and the density of the lower polariton.

Figure~\ref{f:main} shows profiles of the photon wave function $\psi_{\rm
C}$ at $t = 400$ ps.
In Fig.~\ref{f:main} (b), quantized vortex dipoles are generated behind
the obstacle potential.
After the vortex dipoles are released from the potential, they alternately
align in the wake, as shown in Fig.~\ref{f:main} (b).
Such alternate alignment of vortex dipoles in superfluids is also observed
in Refs.~\onlinecite{Nore,Sasaki}.
Figure~\ref{f:main} (c) shows the situation for a larger obstacle
potential.
In this case, two vortices with the same circulation form a pair and
clockwise and counterclockwise pairs are shed alternately.
This vortex shedding dynamics is very similar to that in
Ref.~\onlinecite{Sasaki}, which is a superfluid analogue of the BvK vortex
street.
This periodic vortex configuration has a very long lifetime~\cite{Sasaki}
when there is no dissipation or losses.
Thus, a small obstacle generates vortex dipoles and a large obstacle
generates a BvK vortex street, which is consistent with the case of an
atomic BEC (Fig.~3 of Ref.~\onlinecite{Sasaki}).
For a larger obstacle potential, vortex shedding becomes irregular as
shown in Fig.~\ref{f:main} (d).
The flow velocity in Figs.~\ref{f:main} (a)-\ref{f:main} (d) is $\hbar
k_{\rm p} / m_{\rm LP} \simeq 0.4 v_{\rm s}$, which is comparable to the
critical velocity for vortex nucleation obtained from the
Gross--Pitaevskii equation.~\cite{Frisch}
Figure~\ref{f:main} (e) shows the case for faster flow with flow velocity
of $\simeq 0.7 v_{\rm s}$.
Dark solitons and vortex dipoles are generated in the wake, as
experimentally observed.~\cite{Grosso}
The vortex patterns shown in Fig.~\ref{f:main} can be observed for a
shorter polariton lifetime (e.g., 15 ps) if the polariton density near the
obstacle is kept large by increasing the pump intensity or narrowing the
space between the triangular pumped regions (data not shown).
For all the cases in Fig.~\ref{f:main}, the exciton wave function
$\psi_{\rm X}$ has a similar profile to that of the photon wave function
$\psi_{\rm C}$.

\begin{figure}[htbp]
\includegraphics[width=8cm]{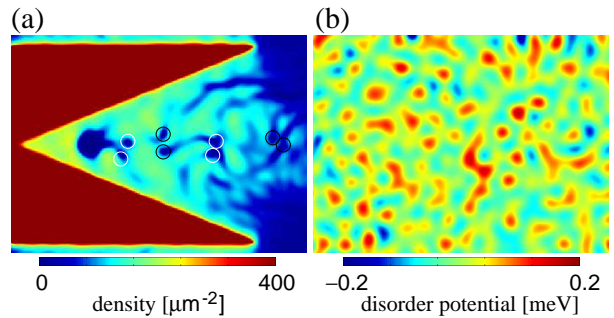}
\caption{
(Color online) (a) Density profile $|\psi_{\rm C}|^2$ in the presence of
the disorder potential shown in (b).
The disorder potential is generated by setting random numbers on each site
and cutting off short-wavelength Fourier components.
The parameters are the same as those in Fig.~\ref{f:main} (c).
The black and white circles indicate clockwise and counterclockwise
vortices, respectively.
See the Supplemental Material for a movie of the dynamics.
}
\label{f:disorder}
\end{figure}
In microcavity samples used in experiments, a disorder potential is
inevitable; it is of the order of 0.1 meV, even for a good
sample.~\cite{Roumpos}
Figure~\ref{f:disorder} shows the effect of the disorder potential on
vortex street formation, where the potential shown in
Fig.~\ref{f:disorder} (b) is added to $V(\bm{r})$ in Eq.~(\ref{GP}) while
the other parameters remain the same as those in Fig.~\ref{f:main} (c).
Figure~\ref{f:disorder} (a) shows that the BvK vortex street is robust
against a disorder potential in a realistic sample.

\begin{figure}[htbp]
\includegraphics[width=8cm]{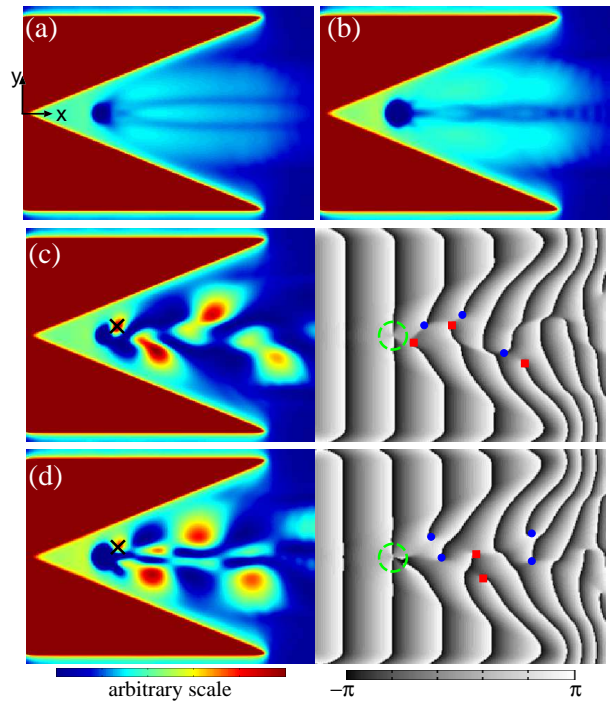}
\caption{
(Color online) (a) Time-integrated density of the photon wave function
$I(\bm{r}) = \int |\psi_{\rm C}(\bm{r}, t)|^2 dt$ for the dynamics in
Fig.~\ref{f:main} (b).
(b) $I$ for the dynamics in Fig.~\ref{f:main} (c).
(c) Density $|G|^2$ (left) and phase ${\rm arg}(G)$ (right) of the
first-order coherence $G(\bm{r}) = \int \psi_{\rm C}^*(\bm{r}_{\rm ref},
t) \psi_{\rm C}(\bm{r}, t) dt$ for the dynamics in Fig.~\ref{f:main} (b).
(d) $|G|^2$ (left) and ${\rm arg}(G)$ (right) for the dynamics in
Fig.~\ref{f:main} (c).
The time integrations in $I$ and $G$ are taken between $t = 400$ and
1400 ps (the patterns are independent of the upper integration limit if
the time integration is sufficiently long).
The position of the reference is $(x_{\rm ref}, y_{\rm ref}) = (46, 4)$,
which is marked by the crosses.
The field of view of each panel is $150 \times 110$ $\mu {\rm m}$ and the
origin is located at the center of the left edge of the panel.
}
\label{f:detect}
\end{figure}
In the present situation, it is impossible to perform time-resolved
measurements, as were performed in Refs.~\onlinecite{Amo09,Nardin,Grosso}
For time-resolved measurements, polaritons must be pumped by a pulsed
laser beam, which is split and used as a reference in the interferometer
with a delay time, enabling time-resolved images to be obtained by
performing repeated measurements.
On the other hand, polaritons are pumped by a cw laser in the present
system and hence only time-integrated images can be obtained.
Figures~\ref{f:detect} (a) and \ref{f:detect} (b) show time-integrated
density profiles $\int |\psi_{\rm C}(\bm{r}, t)|^2 dt$ of the dynamics in
Figs.~\ref{f:main} (b) and \ref{f:main} (c), respectively.
Although the traces of vortex flow are visible, individual vortices are
smeared out and cannot be discerned.~\cite{Amo11}
To overcome this problem, we examine the time-integrated spatial coherence
given by
\begin{equation} \label{G}
G(\bm{r}) = \int \psi_{\rm C}^*(\bm{r}_{\rm ref}, t) \psi_{\rm C}(\bm{r},
t) dt,
\end{equation}
where $\bm{r}_{\rm ref}$ is the position of the reference light source.
The reference light from a small area at $\bm{r}_{\rm ref}$ is enlarged
and interference with the whole image is measured using an interferometer
with a variable arm length.~\cite{Amo11}

Figures~\ref{f:detect} (c) and \ref{f:detect} (d) show the density and
phase profiles of $G$ for the dynamics in Figs.~\ref{f:main} (b) and
\ref{f:main} (c), respectively.
The reference $\bm{r}_{\rm ref}$ is positioned obliquely behind the
obstacle (crosses in Fig.~\ref{f:detect}), where the density oscillates
with time due to periodic vortex shedding.
Therefore, time integration of $G$ is performed stroboscopically at the
same frequency as the change in the vortex pattern, preserving the
information of vortices, as shown in Figs.~\ref{f:detect} (c) and
\ref{f:detect} (d).
Figure~\ref{f:detect} (c), which corresponds to the vortex dipole
generation in Fig.~\ref{f:main} (b), exhibits vortex dipoles in the phase
profiles (left panel of Fig.~\ref{f:detect} (c)).
The phase profile in Fig.~\ref{f:detect} (d) also reflects the BvK vortex
street in Fig.~\ref{f:main} (c).
Thus, the time-integrated coherence $G$ contains information on the vortex
patterns and it enables us to identify vortex street
formation.~\cite{Borgh}

\section{Conclusions}
\label{s:conc}

We have investigated the dynamics of exciton-polariton
superfluid flow passing an obstacle potential in a semiconductor
microcavity.
Numerically solving the two-component nonlinear Schro\"dinger equation in
Eq.~(\ref{GP}) reveals a superfluid BvK vortex street (Fig.~\ref{f:main}
(c)), as well as alternately aligned vortex dipoles (Fig.~\ref{f:main}
(b)) and dark solitons (Fig.~\ref{f:main} (d)).
The formation of the vortex street is robust against a disorder potential,
which is naturally present in samples (Fig.~\ref{f:disorder}).
These periodic vortex patterns are time-dependent steady states
that are attained by cw pumping of polaritons and time-resolved imaging
cannot be performed.
We showed that the spatial coherence in Eq.~(\ref{G}) with an appropriate
reference point $\bm{r}_{\rm ref}$ reflects the vortex patterns even
though it is time integrated (Fig.~\ref{f:detect}).
We have thus demonstrated that a superfluid BvK vortex street can be
generated and detected in an exciton-polariton condensate using current
experimental techniques.

\begin{acknowledgments}
This work was supported by Grants-in-Aid for Scientific
Research (No.\ 22340116 and No.\ 23540464) from the Ministry of Education,
Culture, Sports, Science and Technology of Japan.
\end{acknowledgments}

\end{document}